\documentstyle[aps,preprint,epsfig]{revtex}

\begin{document}


\title{The Thermodynamics of Social Processes: The Teen Birth
Phenomenon}
\author{Nicola Scafetta$^{1}$, Patti Hamilton$^{2}$, Paolo
Grigolini$^{1,3,4}$}
\address{$^{1}$ Center for Nonlinear Science, University of North Texas,\\
P.O. Box 311427, Denton, Texas 76203}
\address{$^{2}$ Center for Nonlinear Science, Texas Womans University, P.O. Box 425498, Denton, Texas 76204}
\address{$^{3}$Istituto di Biofisica CNR, Area della Ricerca di Pisa,
Via Alfieri 1, San Cataldo 56010 Ghezzano-Pisa, Italy }
\address{$^{4}$Dipartimento di Fisica dell'Universit\`{a} di Pisa and
INFM, \\
Piazza\\
Torricelli 2, 56127 Pisa, Italy }
\date{\today}
\maketitle

\begin{abstract}
We argue that a process of social interest is a balance of order and randomness, thereby producing a departure from a stationary diffusion process. The strength of this effect vanishes if the order to randomness intensity ratio vanishes, and this property allows us to reveal, although in an indirect way, the existence of a finite order to randomness intensity ratio. We aim at detecting this effect. We
introduce a method of statistical analysis alternative to the 
compression procedures, with which the limitations of the traditional 
Kolmogorov-Sinai approach are bypassed. We prove that this method 
makes it possible for us to build up a memory detector, which signals 
the presence of even very weak memory, provided that this is 
persistent over large time intervals. We apply the analysis to the 
study of the teen birth phenomenon and we find that the unmarried 
teen births are a manifestation of a social process with a memory 
more intense than that of the married teens. We attempt to give a social interpretation of this effect.
\end{abstract}


\section{introduction}
The concept of complexity is an interesting subject of debate for
philosophers as well as for scientists\cite{cilliers}. It is
interesting to notice that the methods and
concepts borrowed from the physics of complexity are already widely
applied to sociology\cite{eve} in spite of the fact that some aspects
of the physics of complexity are not yet totally understood.
We think that complexity means a delicate balance
between randomness and order, and that complexity turns into
simplicity when one of the components, either order or randomness,
prevails.

It is plausible that social systems, as a 
manifestation of organized societies, rest on the balance of order and 
randomness in the same way as living systems do. Thus, we can use 
models adopted to account for biological processes and apply them to 
the study of processes of sociological interest. To make this aspect 
more transparent for the reader, we first illustrate in which sense a 
DNA sequence can be interpreted as a balance of order and randomness, 
thus illustrating the model itself that will be used in this paper.  
Then we shall shed light on the sociological process here under study 
with the help of that model. 

Let us explain why the DNA sequences can 
be regarded as being a balance of order and randomness. First of all, a 
DNA is a sequence of molecules, either purines or pyrimidines. This 
sequence of molecules can be perceived as a sequence of symbols, with 
$+1$ denoting a purine and $-1$ denoting a pyrimidine \cite{maria}. The site of the DNA sequence, where either a purine or a pyrimidine is located, is conceived as a discrete value of ``time", to which we assign integer values, 1,2,….. We can thus imagine that the DNA sequence is the 
generator of a walk. The walker makes a step ahead or backward, 
according to whether at ``time" $n$ the random walker ``sees" $+1$ or 
$-1$, namely if the n-th site of the DNA sequence hosts a purine or a 
pyrimidine. Now, the important question to answer is whether the walk is totally 
random, as in the ordinary Brownian motion, or there is some kind of 
order in the resulting motion. As well known, the random walk can be 
generated by the Bernoulli map \cite{beck}: Given a sequence of $+1$'s  and 
$-1$'s, generated by tossing a coin, it is possible to find a suitable 
initial condition so that the Bernouilli map results exactly in the 
same ``erratic" sequence of symbols. The advocates of the theoretical 
perspective, according to which  living processes imply a delicate 
balance of order and randomness would rule out the possibility that the 
Bernouilli map is an adequate representation of a DNA sequence. They would rather make the conjecture that an intermittent map is a more adequate representation of it.   In fact, the randomness of an 
intermittent map is rare and would result in a sequence of symbols with 
very extended patches containing only $+1$'s or only $-1$'s. This means that if a given site hosts a purine or pyrimidine, the next site with a high probability will host another purine or pyrimidine.

Actually, the 
authors of Ref. \cite{maria} proved that neither of these two pictures is an 
adequate representation of DNA sequences. In the former picture there 
is no balance between order and randomness, since randomness is the 
only player at work. In the second, the balance would imply an excess 
of order. Thus, the authors of Ref. \cite{maria} proposed a model, called 
Copying Mistaken Maps (CMM) which is a compromise between the two 
pictures. According to the CMM model, nature creates  the DNA sequence 
adopting the following prescription. First, two sequences are 
generated, the former using the Bernouilli map and the latter using an 
intermittent map. The DNA sequence is for the time being a sequence of 
empty sites that Nature has to fill with either $+1$'s or $-1$'s. The 
choices of Nature are probabilistic.  For any site of the real 
DNA sequence to build, Nature adopts the symbol of the corresponding 
site from the former sequence with a probability much higher than that of 
adopting the symbol of the corresponding site of the latter sequence. 
As a consequence of this prescription, the DNA sequence looks like very 
similar to a random sequence, but it is not equivalent to it, and 
detecting the weak memory and correlation is a challenge for the 
statistical analysis of apparently random series. This paper aims at 
illustrating a technique of analysis addressing this challenge. 

It was 
more recently established \cite{marco} that the CMM model is totally equivalent 
to one introduced earlier by another research group \cite{buldyrev}. The model 
proposed by Buldyrev \emph{et al.}, albeit totally equivalent to the 
CMM, makes it possible for us to afford a more proper description of 
the kind of memory that we think to be present in processes of social 
interest, and consequently, we devote  some room to its 
illustration. The authors of Ref. \cite{buldyrev} depicted the DNA sequence as 
resulting from random and totally uncorrelated  fluctuations, whose 
values are either $+1$ or $-1$ with probabilities  $P_{+}$ and $P_{-}$, 
respectively. These probabilities are not equal to $1/2$, but slightly 
different from one another. More precisely, $P_{+} = (1 + 
\epsilon(t)/2$ and $P_{-} = (1 + \epsilon(t))/2$, where $\epsilon(t)$ 
denotes a variable that can only get two values, $2\Delta$ and 
$-2\Delta$, with $\Delta << 1$. The symbol $t$ denotes ``time" in the 
sense earlier described: The symbol $t$ alludes to the fact that the 
site number is very large and it is consequently perceived as a 
continuous variable.  The residence time in one of these two states has 
an inverse power law and the fast random process makes a very large 
number of fluctuations throughout. The paper of Ref. \cite{maria} describes the 
traditional methods adopted in literature to detect this kind of long 
memory, 1)  diffusion analysis, 2)  detrended  fluctuation 
analysis, 3)  Hurst analysis and 4)  spectral analysis. Here we 
propose a new kind of analysis that is applied to the observation of 
the teen birth, the social phenomenon here under discussion.

In an earlier pubblication\cite{west}, the teen birth phenomenon was studied using relative dispersion. Here we analyze 
detrended data for two groups and where this periodic background has been eliminated, 
with the consequent effect of producing two new time series which are 
apparently indistinguishable from random fluctuations. We analyze the 
detrended time series with a new method that has the surprising 
property of detecting a residual memory that cannot be related to 
strong periodic properties, of seasonal nature, which have been 
removed. We term the method proposed in this paper \emph{diffusion 
entropy method}. In the following discussion  we shall explain the reasons for 
this name.

The method  proposed here is of entropic nature and follows the lines 
established many years ago by Kolmogorov\cite{kolmogorov} and
Sinai\cite{sinai}. Their method is referred to as Kolmogorov-Sinai 
(KS) entropic procedure. The traditional KS entropy serves the basic 
purpose of establishing in an objective way if a process is random. 
To detect randomness in a sequence characterized by long-range 
correlation it has been recently proposed\cite{marcoluigi} to replace 
the Shannon indicator with the non-extensive Tsallis 
indicator\cite{tsallis}.
However, the computational detection of the Kolmogorov-Sinai-Tsallis 
(KST) entropy meets the same computational difficulties as that of the 
ordinary KS entropy. Let us explain this in more detail. The DNA 
sequence, according to our prescription of assigning the symbol $\xi 
= 1$ to the purines  and the symbol $\xi = -1$ to the pyrimidines, 
looks like a very large sequence of 
$1$'s or $-1$'. The KS entropy is determined fixing a given window 
of size $N$ and counting the number of times a given combinations of 
$1$' and $-1$' can be accommodated in a window of size $N$. Then, 
we evaluate the corresponding Shannon entropy and we try to assess 
numerically if this quantity divided by $N$ reaches a finite value 
for $N\rightarrow \infty$. The same approach has to be adopted with 
the KST entropy\cite{marcoluigi}, the only significant difference 
being that the Shannon entropy is replaced by the Tsallis entropy.
It is evident that this way of proceeding is limited to windows of
small size, since the number of symbol combinations to take into
account grows  quickly with the window size and rapidly reaches the
maximum size compatible
with the current generation of computers. For this reason an alternate approach involving 
 compression procedures\cite{li,giulia} proves to be an efficient way to bypass 
these limitations.
Here we adopt a different approach which combines the entropic method 
with the diffusion method of Ref.\cite{maria}. In other words, we 
study the non-extensive entropy of the diffusion process generated by 
the statistical data under study according to the prescriptions of 
Ref.
\cite{maria}.

The outline of the present paper is as follows. In Secs. II and 
III, we shall provide the theoretical background for the statistical 
analysis that will be illustrated in Sec. IV. In Sec. III we 
shall apply our method, illustrated in Sec. II, to an artificial 
sequence that is supposed to be a mixture of order and randomness. 
This will contribute an intuitive interpretation of the results of 
Sec. IV and will allow us to draw some conclusions of social 
interest. Section V is devoted to conclusions drawn from the main 
results of this paper.

\section{diffusion entropy}
A process of diffusion in the stationary condition is expected to 
fulfill the important rescaling property
\begin{equation}
p(x,t) = \frac{F(\frac{x}{t^{\delta_{0}}})}{t^{\delta_{0}}}.
\label{stationarycondition}
\end{equation}
Note that $x$ denotes the diffusing variable and $p(x,t)$ is the 
distribution density of this diffusing variable at time $t$.
The nature of the diffusing variable depends on the
complex problem under study. In the case of DNA
sequences\cite{marco,buldyrev} each sequence site is given a value 
$\xi$ either equal to $+1$ or to $-1$ according to whether this site 
hosts a purine or a pyrimidine. The site label $n$ plays the role of 
discrete time, the value $\xi(n)$ is perceived as a time dependent 
fluctuation and the diffusing variable at time $N$ is defined by 
$x(N) \equiv
\sum_{i= 1}^{N}\xi(n)$. In the case of teen births here under study
the variable $\xi(n)$ is a dichotomous sequence given by Eq.(\ref{newsetdata})
and the time $n$ refers to days. Since in both cases $n>>1$ we shall
identify $n$ with the continuous variable $t$.

The diffusion entropy is defined by
\begin{equation}
H(t) = - \int_{-\infty}^{+\infty} dx p(x,t) \ln(p(x,t)).
\label{diffusionentropy}
\end{equation}
Using the stationary condition of Eq.(\ref{stationarycondition}) we
obtain
\begin{equation}
H(t) = A + \delta_{0} \ln(t),
\label{linearincrease}
\end{equation}
where
\begin{equation}
A = \frac{1}{2} \left(1 + \ln(2 \pi \sigma^{2})\right).
\label{A}
\end{equation}
This expression of $A$ is obtained by assuming that $F(y)$ has the
following Gaussian form
\begin{equation}
F(y) = \frac{exp\left(-\frac{y^{2}}{2 \sigma^{2}}\right)}{\sqrt{2\pi
\sigma^{2}}},
\label{gaussianform}
\end{equation}
which defines the meaning of the parameter $\sigma$.

We note that the entropy time evolution of a stationary diffusion
process is linear with respect to logarithmic time
\begin{equation}
\tau \equiv \ln(t),
\label{logtime}
\end{equation}
which makes Eq.(\ref{linearincrease}) read
\begin{equation}
H(\tau) = A + \delta_{0} \tau  .
\label{genericexpression}
\end{equation}

Note that the rescaling index can be, in principle, any real number of the
interval $[0,1]$ and $F(y)$ is not necessarily a Gaussian process.
Moving from  Gaussian to non-Gaussian statistics has  the
effect of changing the analytical expression of $A$ of Eq.(\ref{A}),
but it does not affect the structure of Eq.(\ref{genericexpression}),
provided that the stationary property of
Eq.(\ref{stationarycondition}) holds true.

This paper is devoted to the study of a process of social interest
which is proved to break the stationary condition of
Eq.(\ref{stationarycondition}). In later Sections we shall discuss
the possible origin of this effect for teen births. For the time being, we shall adopt a simplified model, breaking
the stationary condition, which makes
the method of analysis adopted here transparent. The
model is inspired by the earlier work on DNA
sequences\cite{marco,buldyrev}. According to the conclusions of these
papers a DNA sequence can be thought of as a process with fast
fluctuations about a slowly changing bias. It was shown\cite{marco}
that this results in the linear superposition of an  process
of ordinary Brownian diffusion and of L\'{e}vy diffusion. The
statistical weight of the ordinary diffusion component is much larger
than that of anomalous diffusion. Thus, at short times ordinary
diffusion is predominant. Since the component with anomalous diffusion
rests on a larger rescaling parameter $\delta_{0}$ (see
Eq.(\ref{stationarycondition}) for the definition of $\delta_{0}$) at
larger times the anomalous component becomes predominant.
A rough mathematical picture can be obtained expressing the second
moment of the diffusion process, $<x^{2}(t)>$, as follows:
\begin{equation}
<x^{2}(t)> = p_{0}t^{2\delta_{0}} + p_{1} t^{2\delta_{1}}
\label{secondmoment}
\end{equation}
with
\begin{equation}
p_{0} >> p_{1}
\label{strongerrandomness}
\end{equation}
and
\begin{equation}
\delta_{0} < \delta_{1}.
\label{anomalousfaster}
\end{equation}
The transition from ordinary to anomalous diffusion takes place when
the first term on the right hand side of Eq.(\ref{secondmoment}) becomes of the
same order as the second. This takes place at the logarithmic time $\tau = \tau_{ad}$ defined by
\begin{equation}
\tau_{ad} \equiv  \frac{ln(p_{0}/p_{1})}{2(\delta_{1}-\delta_{0}}.    
\label{transition}
\end{equation}

We simulate this form of breakdown of the stationary condition with
\begin{equation}
p(x,t) = \frac{F(\frac{x}{t^{\delta(t)}})}{t^{\delta(t)}},
\label{nonstationarycondition}
\end{equation}
where
\begin{equation}
\delta(t) = \delta_{0} + \eta \ln(t).
\label{logarithmicchange}
\end{equation}
We assume that $\delta_{0} = 0.5$ . Note that the parameter $\eta$ is subtly related to $\tau_{ad}$, and it is expected to become larger with the time $\tau_{ad}$ becoming shorter.  Since the rescaling
parameter $\delta$ cannot exceed the ballistic
value $\delta = 1$, this condition applies to the time scale defined
by
\begin{equation}
\eta \ln(t) < 0.5   .
\label{limitingcondition}
\end{equation}

At this stage it easy for us to show the convenience of adopting the
non-extensive entropy indicator advocated some years ago by
Tsallis\cite{tsallis}. First of all, we notice that in the new
non-stationary condition the traditional entropy indicator yields:
\begin{equation}
H(\tau) = A + \delta_{0} \tau + \eta \tau^{2}.
\label{quadratic}
\end{equation}
Preliminary results on DNA sequences\cite{grigolini} prove that
Eq.(\ref{quadratic}) fits the experimental data very accurately. In
Sec. III we shall show that the theoretical prediction fits the data on teen births  very
well.
This
means that the diffusion entropy has an increase which is faster than the
time increase linear in $\tau$.

The authors of Ref.\cite{marcoluigi}
applied the non-extensive entropy to the analysis of a symbolic sequence
with long-range correlation. The sequence examined in Ref.
\cite{marcoluigi} is equivalent to the CMM
model with a vanishing weight for the random component. In this case
it is shown that the entropy undergoes a regime of linear increase in
time (the time $t$ rather than the logarithmic time $\tau$) if an
entropic index $Q \neq 1$ is adopted. For values of $q < Q$ the
entropy $H_{q}(t)$ undergoes an increase with time faster than the
regime of linear increase in time, and for $q > Q$ the entropy
increase is slower. On the basis of this result we make the
conjecture that in the diffusion regime a linear
dependence on $\tau$ is recovered if an entropic index $q > 1$ is
adopted.

Let us see all this in detail. The non-extensive Tsallis
indicator\cite{tsallis} reads
\begin{equation}
H_{q}(t) = \frac{1 - \int_{-\infty}^{+\infty}dx p(x,t)^{q}}{q-1}.
\label{nonextensiveentropy}
\end{equation}
It is straightforward to prove that this entropic indicator coincides
with that of Eq.(\ref{diffusionentropy}) when the entropic index $q$
gets the ordinary value $q = 1$. Let us make the assumption that in 
the diffusion regime the
departure from this traditional value is very weak. This can be
quantitatively expressed as follows. Let us define first
\begin{equation}
\epsilon \equiv q - 1.
\label{weakdeviation}
\end{equation}
We make the assumption that
\begin{equation}
\epsilon << 1,
\label{conditionofweakdeviation}
\end{equation}
which, as we shall see, is fulfilled by the process under study in this
paper.
This allows us to use the following approximated expression for the
non-extensive entropy

\begin{eqnarray}
H_{1+\epsilon}(t)        &    =     &    - \int_{-\infty}^{+\infty} 
dx p(x,t) \ln(p(x,t))      \nonumber\\
         &        &     -  \frac{\epsilon}{2}  \int_{-\infty}^{+\infty} dx p(x,t) 
\left[\ln(p(x,t))\right]^{2}.
\label{firstorder}
\end{eqnarray}


In the specific case where the non stationary condition of
Eq.(\ref{nonstationarycondition}) applies, this entropy gets the form

\begin{eqnarray}
H_{1+\epsilon} & = & A + \delta(t) \ln(t)  \nonumber\\
  & & - \epsilon B - \epsilon \delta(t) \ln(t) A - \frac{\epsilon}{2} 
\delta(t)^{2}\left(\ln(t)\right)^{2},
\label{explicit}
\end{eqnarray}
where


\begin{equation}
B \equiv \frac{3}{8} + \frac{1}{4} \ln(2 \pi \sigma^{2})
+ \frac{1}{8} \left(\ln(2 \pi \sigma^{2}) \right)^{{2}}.
\label{newdefinition}
\end{equation}
It is straighforward to show that the regime of linear increase in
time is recovered when $\epsilon$ is assigned the value
\begin{equation}
\epsilon = \frac{\eta}{\delta_{0}^{2}/2  + \eta A}.
\label{crucialvalue}
\end{equation}

These theoretical remarks demonstrate that this non-extensive approach to
the diffusion entropy makes it possible to detect the strength of the
deviation from the stationary condition. In fact
Eq.(\ref{crucialvalue}) proves that $\epsilon = 0$ implies a
stationary condition. It is evident that the measure of the departure
from the stationary condition is given by
\begin{equation}
\eta = \frac{1}{2} \frac{\epsilon \delta_{0}^{2}}{1 - \epsilon A}.
\label{nonstationarystrength}
\end{equation}

The conclusion of this Section is that the breakdown of the stationary
property  of Eq.(\ref{stationarycondition}) can be revealed by the
diffusion entropy method under the form of an entropic index $q$
slightly departing from the condition of ordinary statistical
mechanics, namely $q = 1$.

\section{Non-stationary condition induced by weak and persistent 
memory: A simple model}

In this section we show that a sequence of data generated by fast 
fluctuations around a weak and slowly fluctuating drift, produces the 
same effects as those illustrated in Sec. II.
This supports our conviction that the  breakdown of 
the stationary condition discussed earlier 
is a manifestation of weak but persistent memory. In fact, the 
connection between $\eta$ and $\epsilon$ is proven to be the same
as that of Eq.(\ref{crucialvalue}) and
Eq.(\ref{nonstationarystrength}). This means that for short periods of 
the logarithmic time $\tau$ the effect of persistent memory becomes 
indistinguishable from the breakdown of the stationary condition.
We have in mind the notable effect, illustrated in Sec. II, of the 
structure of
of  Eq.(\ref{logarithmicchange}) yielding the entropic property of 
Eq.(\ref{quadratic}).

We create a sequence of data, $\xi_{\Delta}(t)$, where $\Delta$ 
denotes the memory intensity, in the following way. Firstly, using a 
random noise generator we create a sequence of fluctuations. More 
specifically, we generate a set of 100,000 random rational numbers, 
$F(n)$, belonging to the interval $[0,1]$. The variable $n$ is an 
integer number running from 1 to 100,000.
Secondly, we create an artificial memory through a square periodic 
function, $f_{\Delta}(t)$, with period equal to 2000, average equal 
to 0.5, and amplitude equal to the parameter $\Delta$, which is, as 
 mentioned earlier, the memory intensity. In the first period, for $t$ 
from 0 to 2000, the function $f_{\Delta}(t)$ is:

\begin{equation}
f_{\Delta}(t) = \left\{
\begin {array} {rrrl}
0.5 + \Delta \quad& if & 0 < t \leq 1000 &  \\
0.5 - \Delta \quad& if & 1000 < t \leq 2000 & .
\end{array}  \right.
\label{simulfunction}
\end{equation}

In the third and final stage we convert the sequence $F(n)$  into the 
dichotomous
sequence, $\xi_{\Delta}(t)$, of numbers ``+1" and ``-1" using the 
following prescription:

\begin{equation}
\xi_{\Delta}(t) = \left\{
\begin {array} {rrrl}
+1 \quad& if & F(t) > f_{\Delta}(t) & \\
-1 \quad& if & F(t) < f_{\Delta}(t) & .
\end{array} \right.
\label{simulation}
\end{equation}

The numbers ``-1" and ``1" can be interpreted as the discrete jumps 
of a random walker. This means that with the data $\xi_{\Delta}(t)$, 
interpreted as the steps made, forward or backward, by a random 
walker moving on the $x$-axis, we can build a trajectory, denoted by 
$\Xi_{\Delta}$. This trajectory specifies the random walker's position 
on the same $x$-axis, at any given time $N$.
With $\Delta=0$ there is no memory: The data $\xi_{\Delta=0}(t)$ are 
statistically equivalent to those that one would obtain by tossing a 
fair coin.
This model is similar to the CMM model described in Sec. 
I, the only significant difference being that the time duration of the 
states $\Delta$ and $-\Delta$ here is fixed, and equal to $1000$, 
whereas in the CMM the distribution of these time durations is an 
inverse power law. We expect however, that the effect here detected is 
very close to that produced by the CMM when the mean time duration is 
equal to $1000$.

To apply the diffusion entropy method we need to produce diffusion 
first, and this implies the use of a very large number of 
\emph{auxiliary trajectories}. These trajectories are created from 
the \emph{original 
trajectory,}$\Xi_{\Delta}$. Let us see in detail how we create the original and the 
auxiliary trajectories.
The original trajectory is obtained as follows. The random walker 
starts from the initial position assumed to be $t = 0$. At the time 
$t=1$ it occupies the position corresponding to the first jump, 
$\xi_{\Delta}(1)$. At time
$t=2$ the random walker will occupy the position $ \Xi_{\Delta}(2) = 
\xi_{\Delta}(1)+ \xi_{\Delta}(2)$, and, more in general at the time 
N, will occupy the position $ \Xi_{\Delta}(N) = \xi_{\Delta}(1)+ 
\xi_{\Delta}(2)+...+\xi_{\Delta}(N)$.
The auxiliary trajectories are built up
in the same way by assigning the role of first jump to 
$\xi_{\Delta}(r)$, with $r=1,2,3,\ldots$.
  Formally we define the $r^{th}$ trajectory as:
\begin{equation}
x_{\Delta}(r, t) \equiv
\sum_{i= 0}^{t}\xi_{\Delta}(i+r),
\label{manyfromone2}
\end{equation}
with $r = 1, 2, 3,\ldots$.
The values $r=1,2,3,\ldots$ determine the auxiliary trajectories. Of 
course, the number of trajectories
is determined by the number of data at our disposal. For example, 
with the 100,000 data used in the simulation of this section, it is 
possible to obtain 99,900 trajectories of 100 jumps each.

At this point, it is possible to calculate the diffusion entropy 
associated with the weak memory controlled by the parameter $\Delta.$ 
First, we use the trajectories obtained with the earlier prescription 
to calculate the probability $p_{\Delta}(x_{\Delta},t)$ of finding 
the random walker in the position $x_{\Delta}$ after $t$ jumps. 
Second, we calculate the diffusion entropy, $H_{\Delta}(t)$, at the 
$t^{th}$ jump with the following formula:

\begin{equation}
H_{\Delta}(t) = -  \sum _{ x_{\Delta}}  p_{\Delta}(x_{\Delta},t) \log 
(p_{\Delta}(x_{\Delta},t)),
\label{entropysum}
\end{equation}where the sum is understood on all available positions 
$x_{\Delta}$.

We illustrate in Fig. 1 the results of this numerical analysis. We 
plot three curves corresponding to $\Delta=0,$  $\Delta=0.04,$ and 
$\Delta=0.10.$ For $\Delta=0$ the diffusion entropy refers to a 
stationary diffusion process. According to the theory of Sec. II, 
and more precisely, according to Eq.(\ref{genericexpression}), this 
condition is expected to produce an entropy increase linear with 
respect to the logarithmic time $\tau=\ln(t)$. In the case under 
study, the stationary condition is that of an ordinary Brownian 
diffusion, which yields for the rescaling coefficient $\delta_{0}$ 
the value $0.5$. Thus, the diffusion entropy is given by

\begin{equation}
H(t) = 0.5\ln(t) + \ln(2),
\label{randomwalk}
\end{equation}
which fits very well the numerical result of Fig. 1.
For the other two curves, corresponding to
$\Delta=0.04$ and $\Delta=0.10$, respectively, there is a significant 
deviation from the linear dependence on the logarithmic time $\tau$, 
which is larger with the larger memory strength. In conclusion, Fig.1 
shows that the numerical evaluation of the diffusion entropy detects 
the breakdown of the condition of stationary diffusion, even if this 
is caused by a very weak memory. In the case under study, the memory 
strength, given by the parameter $\Delta$, is equivalent respectively 
to the 4\% and 10\% of the signal.

Let us discuss now how to measure the intensity of the breakdown of 
the stationary condition. In accordance to the prescriptions of 
Sec. II, this can be done in two different ways. The former method is  
direct. It is based on fitting the curves with the quadratic 
approximation of Eq.(\ref{quadratic}).
This allows us to determine the coefficients $A,$ $\delta_0,$ and 
$\eta$. On the basis of the theoretical remarks of Sec. II we realize that the latter parameter is the property of interest to measure. We can 
establish a connection with the second method by determining the 
entropic index $q=1+\epsilon$ by means of Eq.(\ref{crucialvalue}).
The latter method rests on the direct determination of the entropic 
index $q$, as the ``magic" value of $q$ making the non-extensive 
Tsallis entropy linear with respect to logarithmic time $\tau$. In 
practice, this means that we have to look for the value of $q$ that 
results in the maximum the coefficient of linear correlation. The form of the 
non-extensive Tsallis entropy used in this case is:

\begin{equation}
H_{q, \Delta}(t) = \left( 1-\sum 
_{x_{\Delta}}p_{\Delta}(x_{\Delta},t)^q\right)/\left(q-1 \right),
\label{tsallisentropy}
\end{equation}
  where, as in the earlier case, the sum is understood on all 
available positions $x_{\Delta}$. For $q=1$ Eq.(\ref{tsallisentropy}) 
is identical to the Eq.(\ref{entropysum}).

We illustrate the results of the numerical calculations based on the 
adoption of Eq.(\ref{tsallisentropy}) in Fig.2. This figure
refers to the same physical conditions as those of Fig.1 and proves 
that for any of those conditions an entropic index $q$ can be found 
so that the non-extensive Tsallis entropy becomes a linear function 
of $\tau$. Thus, within the context of the problems under discussion 
in this paper,  Tsallis entropy takes on the following 
"thermodynamic" meaning. This new type of entropic indicator  allows us 
to imagine the processes departing from the 
stationary condition as also being stationary. The departure of $q$ from the ordinary value 
$q=1$ increases with increasing $\delta$. We find that $\Delta=0$, 
$\Delta=0.04$ and $\Delta=0.10$ correspond to $q=1$, $q=1.054$ and 
$q=1.205.$, respectively.
All this is in complete accordance with the theoretical remarks of Sec. II.

We make a judgment of the results obtained by means of the numerical 
treatment of the simple model of this section with the help of Table 
I. In the first three columns of this table we report, for different 
values of the parameter $\Delta$, the values of the coefficients $A$, 
$\delta_0$, $\eta$ and $q=1+\epsilon$ calculated using the 
Eq.(\ref{crucialvalue}), and thus using the former of the two methods 
 illustrated earlier. In the last column of this table we report the 
values of $q$ obtained using the latter method. The fitting procedure 
adopted to generate
the values of this table have been limited to time windows whose size 
does not exceed the value of $30$. We see that, as expected, $q$ is 
an increasing function of
$\Delta$ and that within the statistical accuracy the two methods 
yield the same value for $q$. This means that Eqs.(\ref{crucialvalue}) 
and(\ref{nonstationarystrength})are correct.
Of course, this implies that the memory strength to be weak enough. 
 From the values reported in Table I we see that the accuracy of the 
theoretical prediction is satisfactory for values $\epsilon \leq 
0.207$, a fact which implies the maximum value $\Delta=0.09$ for the 
memory strength. Beyond this value the validity of the quadratic 
approximation necessary to evaluate the parameter $\eta$ is broken. 
For higher values of the memory strength it is probably convenient to 
use the latter method. However, in this case the theoretical 
connection with the breakdown of stationary diffusion is still 
missing and further research work is required.

\section{Application to the teen birth phenomenon}

In this section we give a brief introduction to the teen birth phenomenon in Texas and place the results of the diffusion entropy method of analysis within the context of earlier research. The reader will find more detailed discussion of analysis of teen births in Texas in an earlier publication\cite{west}. then we show the method of diffusion entropy at work.

\subsection{The teen birth phenomenon}

Texas is second only to California in the number of births to teens in the United States. Rates of birth to teens of all ages and racial/ethnic groups have been dropping since 1990, Ref.\cite{Ventura}. However, the size of the problem in Texas remains significant.

In 1996, in Texas there were 80,490 pregnancies and 52,273 births to girls 15-19 years old, Ref.\cite{national}. The U.S. rate of pregnancy among young women 15 to 19 years old was 97 per 1000 girls of that age, the rate in Texas was 113 per 1000. The mean age of teens giving birth was 17.62 years in Texas. Approximately 66\% of teen births in Texas were out of wedlock and 24\% of births to teens were to girls who had given birth at least once previously.

Data used in Ref.\cite{west} to study the nonlinear dynamics in teen birth data included daily counts of all births to teens in Texas from 1980 through 1998. Findings demonstrated the teen birth data obeyed a scaling law. The authors\cite{west} concluded the scaling relation tied together what happened at the shortest time scale with what happened at the longest time scale, thus, resulting in long-term memory. When found, such long term correlation and complexity in time series suggests there is strong feedback across time scales in the process. These authors\cite{west} suggest the phenomenon is dominated by a self-induced stability which may increase with population density, mobility and interaction among persons, between persons and social institutions, and among social institutions. To date these conjectures have not been tested.

The authors of Ref.\cite{west} did not detrend or smooth the data prior to their analysis, nor did they investigate the effects of marital status on the scaling process. It was their intent to study the gross behavior of the time series of all births. The approach used here provides a deeper look into the endurance of memory and its possible source for this data.

Data for the study reported here were abstracted from birth certificates obtained from the Texas Department of Health. The original time series was constructed from the daily count of births from January 1, 1994 through December 31, 1998. Every recorded birth to a woman under the age of 20 was included. Data on the marital status of the mother allowed us to analyze married and unmarried births separately. Reliable and valid birth certificate information regarding marital status did not become available in Texas until January 1, 1994. See Table II for further description of the data.

The reason marital status is relevant to the analysis that follows is based on the observation that amount of linear memory differs between time series of births to married and to unmarried teens. See Table III for autocorrelations for lags of one weak and approximately one year in total teen births and married and umarried teen births separately. The fewest births to both married and unmarried teens occurred in the second quarter of each year (April, May and June). The third quarter of every year (July, August and September) had significantly more births to teens than any other quarter.

Large third quarter peaks in births and autocorrelation = .536 at lags of 364 days were found for all teen births. Atmospheric factors such as light and temperature have been suggested as reasons for annual cycles in human births, Ref.\cite{lammiron}. However, it is not clear why such factors would have a stronger effect on unmarried than on married teens as indicated by differences in autocorrelation at lags of 364 days.

Strong weekly periodicities (autocorrelation = .665 at lags of 7 days for all teens) were found. These may have been imposed on the process through provider preference. Scheduled inductions of labor and cesarean sections generally occur early in the week in order to assure patients are out of the hospital by the weekend when staffing is a greater problem. However, further investigation is needed to determine the veracity of the assumption that these preferences are related to the weekly periodicity observed in the data. In addition, investigation is needed to determine the reason for the disparity in autocorrelation for lags of 7 days between married and unmarried births.

\subsection{Preliminary detrending}

This paper aims to establish a new technique of analysis to discover weak memory remaining after the removal of linear correlations. We are aware that any form of detrending carries with it the risk of loss of information and obliteration of long-range memory. However, in this case we believe detrending provided us an opportunity to test our new method for sensitivity to long-lasting memory of extremely weak intensity. We believe earlier findings\cite{west} are the result of long-lasting memory of large intensity, perhaps the result of annual periodicities not removed by detrending. We make the conjecture that, although weak, there is a residual memory left after detrending annual periodicity, steady drift and removing births occurring on weekends and holidays. In Sec. V we shall try to explain the implications of the results. With these warnings in mind,  we proceed with our detrending prescription as follows.

First of all, we notice that the number of births corresponding 
to Saturdays and Sundays are much smaller than those of the week days. 
In addition to  Saturdays and Sundays, there are other days with very 
small number of births that are identified with holidays. All data 
identified as holidays are erased and replaced by empty sites. The data 
of Fig. 3 report the daily number of births with empty sites that are 
not visible in the scale of that figure. The data refer to the married 
(a) and unmarried (b) teens, in the state of Texas from 1994 to 1998. 
In both cases the data show ostensible signs of seasonal
periodicities. We remind the reader that the purpose of this paper is to prove that the diffusion entropy method is an efficient way of detecting residual memory after detrending trivial periodic properties, thus revealing the balance between order and randomness.
The seasonal recurrences shown in Fig. 3 seem to be a significant 
example of obvious periodic property.

Thus, we detrend it proceeding as 
follows. We assume that the periodic property is described by
\begin{equation}
\Xi(t) = A + B t + C cos (\omega t ) + D sin(\omega t).
\label{fittingcurve}
\end{equation}
This is a deterministic process whose form is determined by seasonal 
periodicity, the harmonic part, and by the fact that the number of 
births is proportional to the increase (or decrease) of teens within 
 the two categories with time, the term $Bt$. In the case of the 
unmarried teens we set A = 97.5, B = 0.00893, C = 1.29, D = -6.30 and 
$\omega = 2 \pi/365.25.$  In the case of the married  we set A = 57.8, B 
= -0.00353, C = - 0.277, D = -4.14 and $\omega = 2 \pi/365.25.$  The 
resulting curves are illustrated by the solid lines of Fig. 3a and Fig. 
3b.  In Fig. 4a and 4b we show the time series after application of 
this detrending procedure. It is evident that the data of Fig. 4 look 
more erratic than the original data of Fig. 3. A mere sight inspection 
of Fig. 4 seems to indicate  that the data concerning unmarried teens 
are more correlated than those of the married. The diffusion entropy 
method of this paper aims at a quantitative assessment of this 
property. To proceed to this quantitative assessment that will be 
illustrated in Sec. IV C, we have to prepare the data in a suitable 
way. We call these new data $\zeta_{a}(n),$ if they refer to unmarried 
teens, and $\zeta_{b},$ if they refer to unmarried teens. The symbol n 
denotes, as usual, the discrete "time" n = 1,2,3.... It is important to 
point out that this time does not correspond exactly to the days of the 
original data, because the empty sites are not counted, and if two 
given days are separated by holidays, the latter day is labelled as 
being immediately subsequent to the former.

\subsection{Diffusion entropy}

We apply the diffusion entropy method to the new data $\zeta_a(n)$ and
$\zeta_b(n)$ of Fig. 4(a) and Fig. (4b)  using the recipe illustrated in Sec. III. To make easier for us to adopt this recipe, we will transform these two sequences into two 
dichotomous sequences. To do that we adopt the following prescription

\begin{equation}
\xi_{a/b}(n) = \left\{
\begin {array} {rrrl}
+1 \quad& if & \zeta_{a/b}(n) > 0 & \\
-1 \quad& if & \zeta_{a/b}(n) < 0 & .
\end{array} \right.
\label{newsetdata}
\end{equation}
As done in Sec. III, we create a large number of trajectories 
described mathematically by

\begin{equation}
x_{a/b}(r,t) \equiv
\sum_{i= 0}^{t}\xi_{a/b}(i+r),
\label{manyfromo}
\end{equation}
with $r = 1, 2, 3,\ldots$. This corresponds  to the prescription of 
Eq.(\ref{manyfromone2}). Then, following again Sec. III, we 
calculate the probability $p_{a/b}(x_{a/b},t)$ of finding find the random 
walker in the position $x_{a/b}$ after $t$ jumps.  Finally, we 
calculate the diffusion entropy of the two sets of data, 
$H_{a/b}(t),$ with the following formula:

\begin{equation}
H_{a/b}(t) = -  \sum _{ x_{a/b}}  p_{a/b}(x_{a/b},t) \log (p_{a/b}(x_{a/b},t)),
\label{entropysu2}
\end{equation}
which corresponds to Eq.(\ref{entropysum}).

We show the results of this statistical analysis in Fig.5. This 
figure shows that the diffusion entropy of both sets of data 
exhibits deviation from the stationary condition, indicated by the 
dotted line. The departure from ordinary statistical mechanics of the 
unmarried teens is much larger than that of the married teens. The 
curves of Fig. 5 are remarkably similar to those of Fig. 1 and this 
fact, by itself, suggests that births to unmarried 
teens contain  more memory than is the case for married teens.

Following the prescriptions of Sec. III, we also use the Tsallis 
entropy, which, in this case, reads:
  \begin{equation}
H_{q, a/b}(t) = \left( 1-\sum 
_{x_{a/b}}p_{a/b}(x_{a/b},t)^q\right)/\left(q-1 \right).
\label{tsallisentropy2}
\end{equation}
We illustrate the results of this analysis in Fig. 6. This figure 
shows that both the diffusion entropy of the unmarried teen and the 
diffusion entropy of the married teens have a linear dependence on 
the logarithmic time $\tau$ if we use for the former case $q = 1.204$ 
and for the latter $q=1.050$.

Finally, as done in Sec. III, in Table IV we compare
the two distinct methods adopted for the detection of the entropic 
index $q$. We see that the results of the latter 
method are very close to those of the former, thereby ensuring the 
validity of the parabolic fitting of the former method. Actually, the 
agreement in the case of the married teens is better than in the case 
of the unmarried teens, which are characterized by memory strength 
larger than that of married teens. This is in agreement with the 
remarks of Sec. III suggesting that the strong memory case can 
produce a conflict between the two methods. However, we think that the hidden 
memory, detected by the new method of this paper, is weak enough as 
to ensure the validity of our main conclusion that the 
married and unmarried cases show a remarkable similarity with the 
memory strengths $\Delta = 0.04$ and $\Delta = 0.10$ of Fig. 2, 
respectively. More precisely, according to Table IV, the married and 
unmarried case correspond to $\eta = 0.006$ and $\eta = 0.035$, 
respectively. We conclude that the method of diffusion entropy 
affords a reliable proof that births to unmarried teens have 
  stronger `weak' memory than those to married teens.

One might wonder if the procedure of making  the signal to 
analyze dichotomous might have produced statistical effects distinct 
from those resulting from the actual signal. Numerical calculations, 
not reported here, prove that it is not so, and that the actual data 
produce similar results. For the quantitative purpose of measuring the 
memory strength the adoption of a dichotomous signal yields the benefit 
of resting only on integer numbers for the production of the histograms 
necessary to the entropy calculations. Furthermore, the adoption of 
dichotomous signals provides a deeper  connection with the DNA model 
described in Sec. I, and it is in fact the reason for the surprising 
similarities between Fig. 1 and Fig. 5 and between Fig. 2 and Fig. 6.

\section{concluding remarks}

As a first interesting result, this paper establishes that 
there is an intimate relation between memory and the breakdown of the 
stationary condition expressed by the rescaling property of Eq. (1). It 
has to be pointed out that the departure of the entropic index $q$ from 
the ordinary value $q = 1$ does not signal a departure from ordinary 
statistics, as claimed in Ref.\cite{tsallis2}. Rather, it signals the 
breakdown of the rescaling property of Eq. (1), which, in turn, is due 
to the fact that the signal under study consists of fast fluctuations 
with a time dependent bias that are a much slower function of time. If 
the bias is weak, the deviation from the ordinary entropic index $q = 
1$ can be related to the memory strength $\eta$, which can be 
independently derived from the adoption of the ordinary Shannon 
entropic indicator. In other words, the parameter $\eta$ is more 
significant than the parameter $\epsilon \equiv q - 1$, since it 
correctly suggests that the effect revealed by the diffusion entropy 
method has to do with the time dependence of the rescaling parameter 
$\delta$. We recover from a different perspective the conclusions of 
earlier work \cite{mauro}. The L\'evy processes are an interesting example of 
non-Gaussian statistics. Yet, their diffusion entropy would yield a 
linear increase with respect to the logarithmic time $\tau \equiv ln(t)$, 
described by Eq. (7), even if in this case $\delta_{0}$ would be a 
rescaling parameter larger than $0.5$. This is so because the dynamical 
derivation of the L\'evy processes, after a process of memory erasure 
\cite{mauro}, yields the rescaling property of Eq. (1).

The second important result of the paper is that the method of 
diffusion entropy is very sensitive to weak but persistent memory. The 
siniliarities between Fig. 1 and Fig. 5 and between Fig. 2 and Fig. 3 
are impressive. These figures mean that it is plausible to conjecture 
that the detrending process used to analyze the birth data does not 
eliminate all forms of memory. The form of memory adopted in Sec. 
III to create the artificial sequence is a deterministic process with 
the time period of 2000 days. This cannot be considered as a proof that 
this is the order of magnitude of the hidden periodicity. We can only conjecture that the periodicity responsible for the deviation of entropy from the linear increase with respect to the logarithmic time $\tau$, is larger than the maximum observation time, which is of the order of 30 days. Beyond 30 days the number of trajectories available is too scarce to ensure statistical stability.

Here we make two conjectures regarding sources of weak memory 
detected in the detrended data. First, it is likely that weekly and 
annual periodicities removed by detrending were not the only 
periodicities present in the data. For example, there are weak cycles with 
periods of approximately one half year remaining in the data for unmarried 
teens after detrending.  Similar cycles do not appear in the data for 
married teens. There are sociological factors such as school schedules 
and holiday breaks that may account for the difference in weak 
sub-annual cycles between married and unmarried teens.  While these 
effects have not yet been investigated, further study could reveal 
their association to the memory strengths of the data.  However, this 
conjecture seems simplistic in light of our second conjecture regarding 
fractal scaling in the data. Traditional time series analysis methods 
used in the social sciences are based on the assumption that 
complicated time series data represent the superposition of numerous 
frequencies of varying periods. A frequency for which cause is 
understood or considered trivial is modeled and/or removed. Frequencies 
are, thus, explained and removed iteratively until all useful 
information in the data is accounted for and the data remaining 
represent white noise. However, the time series data used to test the 
methods described here are known to have fractal scaling properties 
\cite{west}. These properties are the result of feedback across all time scales 
within the time series (days, weeks, months, years).  Fractal scaling 
processes cannot be fully characterized by the superposition of independent, additive 
frequencies. Instead, the data are nonlinear in that all frequencies 
are folded together in a complex pattern. In such a case, detrending, 
as we prescribed, would be insufficient to obliterate the effects of 
weekly and annual periodicities folded into periodicities on other time 
scales. Thus, the weak memory remaining may be the result of fractal 
scaling of the data that resists simple detrending techniques common 
in the social sciences.  The difference in scaling properties of 
married and unmarried teen births has not been thoroughly investigated 
but suggests unmarried teens are affected by a stronger feedback 
mechanism than are married teens.

All these conjectures suggest that the logarithmic oscillations 
detected in Ref. \cite{west} are of such large intensity because of the fact that 
the yearly periodicities are not detrended. Their detrending, as shown 
in this paper, makes the resulting signal much closer to the ordinary 
random walk. The analysis of this paper reveals, however, that this is 
not an ordinary random walk, and that a fractal cascade of frequencies of 
smaller and smaller intensity might exist. This is a challenge for future 
application of the entropic method of analysis developed in this paper. 
We plan to shed light on this and other intriguing issues raised by our method by means of the joint 
analysis of real data and of artificial sequences like that of Sec. 
III, with fluctuating periodicities driven by an inverse power law 
prescription.

\begin{center}
ACKNOWLEDGMENTS
\end{center}
Partial support for this study was obtained from the National Institute 
  for Child Health and Human Development, grant  R03HD37207-02.


\newpage

\begin{center}
TABLE I

\begin{tabular}{||c||c|c|c|c||c||}\hline
\multicolumn{6}{|c|}{\bf  Entropic index q as resulting}\\
\multicolumn{6}{|c|}{\bf  from two distinct fitting procedures }\\ \hline
$\Delta$ &  $\eta$  &  $\delta_0$  &  $A$ & $q=1+\epsilon$ & $q$ \\ 
\hline \hline
   0.03  & 0.0018               &  0.517              &  0.690 
&  1.014           &  1.017  \\
            & $\pm$.0005   & $\pm$.002  & $\pm$.002   & $\pm$..004 
& $\pm$.008    \\ \hline
  0.04  & 0.0063               & 0.512   & 0.692   & 1.047   & 1.054   \\
            &  $\pm$.0006  & $\pm$.002          & $\pm$.002   & 
$\pm$.005   & $\pm$.005   \\ \hline
   0.05  &  0.0113              & 0.506    & 0.693   & 1.083   & 1.092   \\
            & $\pm$.0007   &  $\pm$.003         & $\pm$.003   & 
$\pm$.006   & $\pm$.004   \\ \hline
   0.06  & 0.0165               & 0.501   & 0.695   & 1.120   & 1.127   \\
            & $\pm$.0007   & $\pm$.003   & $\pm$.003   & $\pm$.007   & 
$\pm$.004   \\ \hline
  0.07   &  0.0214              & 0.498   & 0.696   & 1.154   & 1.157   \\
            &  $\pm$.0008  & $\pm$.003   & $\pm$.003   & $\pm$.008   & 
$\pm$.004  \\ \hline
  0.08  &  0.0261                & 0.497   &  0.696  & 1.184   & 1.181   \\
            &  $\pm$.0007  &  $\pm$.003  & $\pm$.00   & $\pm$.008   & 
$\pm$.004   \\ \hline
  0.09  &  0.0300                & 0.498   &  0.695  & 1.207   & 1.197   \\
            &  $\pm$.0006  &  $\pm$.003  & $\pm$.002   & $\pm$.007   & 
$\pm$.004   \\ \hline
  0.10  &  0.0329                & 0.503   &  0.693  & 1.221   & 1.205   \\
            &  $\pm$.0005  &  $\pm$.002  & $\pm$.002   & $\pm$.005   & 
$\pm$.004   \\ \hline
\end{tabular}
\end{center}

\newpage

\begin{center}
TABLE II

\begin{tabular}{|c|c|c|c|}\hline
\multicolumn{4}{|c|}{\bf  Marital Status }\\
\multicolumn{4}{|c|}{\bf  Data Set Used N=1994-1998=1826 days }\\ \hline
Data &  All  &  Married  &  Unmarried  \\
Set  &  Teens  &  Teens  &  Teens  \\ \hline
Mean \# & &  &   \\
  Daily Births& 149.14 & 50.39 &  98.52 \\ \hline
Range & 143 & 68 & 97  \\ \hline
Minimum & 81 &  22 & 55 \\ \hline
Maximum & 224 & 90 & 152  \\ \hline
Standard &  &  &  \\
   Dev  & 23.52 & 10.34 & 16.68  \\ \hline
Variance & 553.03 & 106.89 & 278.15  \\ \hline
Total  & &  &   \\
  Births &272,328* & 92,006 & 179,893  \\ \hline
\end{tabular}

*Marital status was missing on 429 teen birth certificates.
\end{center}

\newpage

\begin{center}
TABLE III

\begin{tabular}{|c|c|c|c|}\hline
\multicolumn{4}{|c|}{\bf  Autocorrelation}\\
\multicolumn{4}{|c|}{\bf  in Married and Unmarried Teens }\\ \hline
Data &  All  &  Married  &  Unmarried  \\
Set  &  Teens  &  Teens  &  Teens  \\ \hline
Lags of & &  &   \\
  7 Days & .665 & .466 &  .602 \\ \hline
Lags of & &  &   \\
364 Days & .536 & .370 & .464  \\ \hline
\end{tabular}
\end{center}

\newpage

\begin{center}
TABLE IV

\begin{tabular}{||c||c|c|c|c||c||}\hline
\multicolumn{6}{|c|}{\bf  Entropic index q as resulting}\\
\multicolumn{6}{|c|}{\bf  from two distinct fitting procedures }\\ \hline

            &  $\eta$  &  $\delta_0$  &  $A$ & $q=1+\epsilon$ & $q$ \\ 
\hline \hline
  unmarried  &  0.035              &  0.513              &  0.687 
&  1.23           &  1.204  \\
  teens       & $\pm$ 0.002     & $\pm$ 0.007    & $\pm$ 0.007   & 
$\pm$ 0.02  &     \\ \hline
  married  &  0.006                  &  0.514               & 0.697 
&  1.046            & 1.050   \\
  teens      &  $\pm$ 0.001      & $\pm$ 0.004      & $\pm$ 0.004   & 
$\pm$ 0.009  &    \\ \hline
\end{tabular}
\end{center}

\newpage

\begin{figure}[h]
\epsfig{file=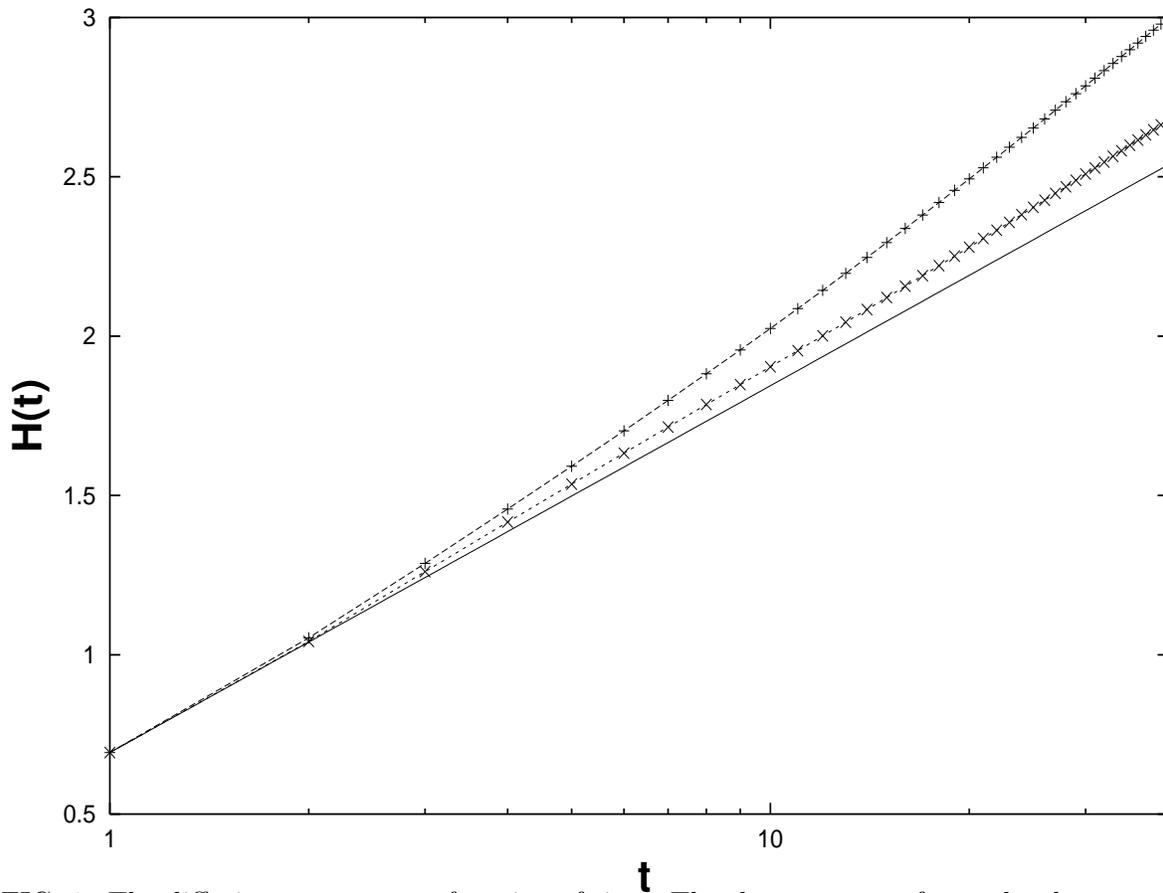, height=16cm,width=12cm,angle=-90}
\caption{The diffusion entropy as a function of time. The three curves
refer  to the three sets of data with $\Delta=0$ (solid line), 
$\Delta=0.04$ (the curve denoted by $\times$) and $\Delta=0.10$ (the 
curve denoted by +), respectively. The case $\Delta=0$ results in the 
diffusion entropy of a stationary process, the ordinary random walk, 
in this case.  The corresponding curve, as expected, is a linear 
function of the logarithmic time $\tau \equiv \ln(t)$, see 
Eq.(\ref{randomwalk}). The other two curves, corresponding to 
non-vanishing memory strength, result in an evident departure from 
the linear dependence on logarithmic time, larger for the case of 
larger memory (larger $\Delta $). This is a clear illustration of the 
breakdown of the stationary condition caused by a memory of weak but 
non-vanishing intensity.}\end{figure}

\newpage

\begin{figure}[h]
\epsfig{file=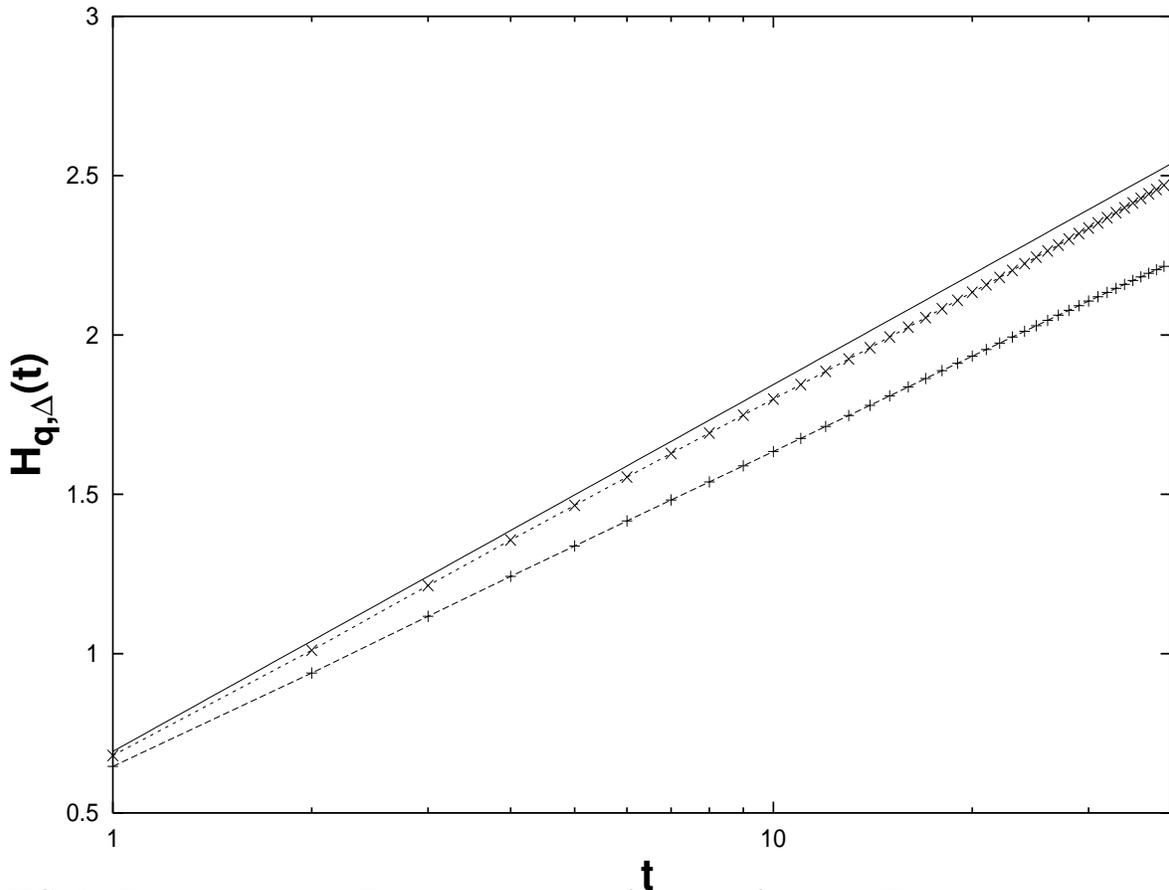, height=16cm,width=12cm,angle=-90}
\caption{The non-extensive Tsallis entropy as a function of time $t$. 
The three curves are the numerical realization of 
Eq.(\ref{tsallisentropy}) with $q=1$ (solid line), $q=1.054$ (symbol 
$\times$) and $q=1.205$ (symbols +) and correspond to different 
values of the memory strength $\Delta$, which are $\Delta=0$, 
$\Delta=0.04$ and $\Delta=0.10$, respectively. The choice of 
different entropic indices $q$ for the different values of $\Delta$ 
has been done with the criterion of selecting
the value of $q$ resulting in the most extended
linear regime with  respect to the logarithmic time $\tau \equiv \ln(t).$ }
\end{figure}

\newpage

\begin{figure}[h]
\epsfig{file=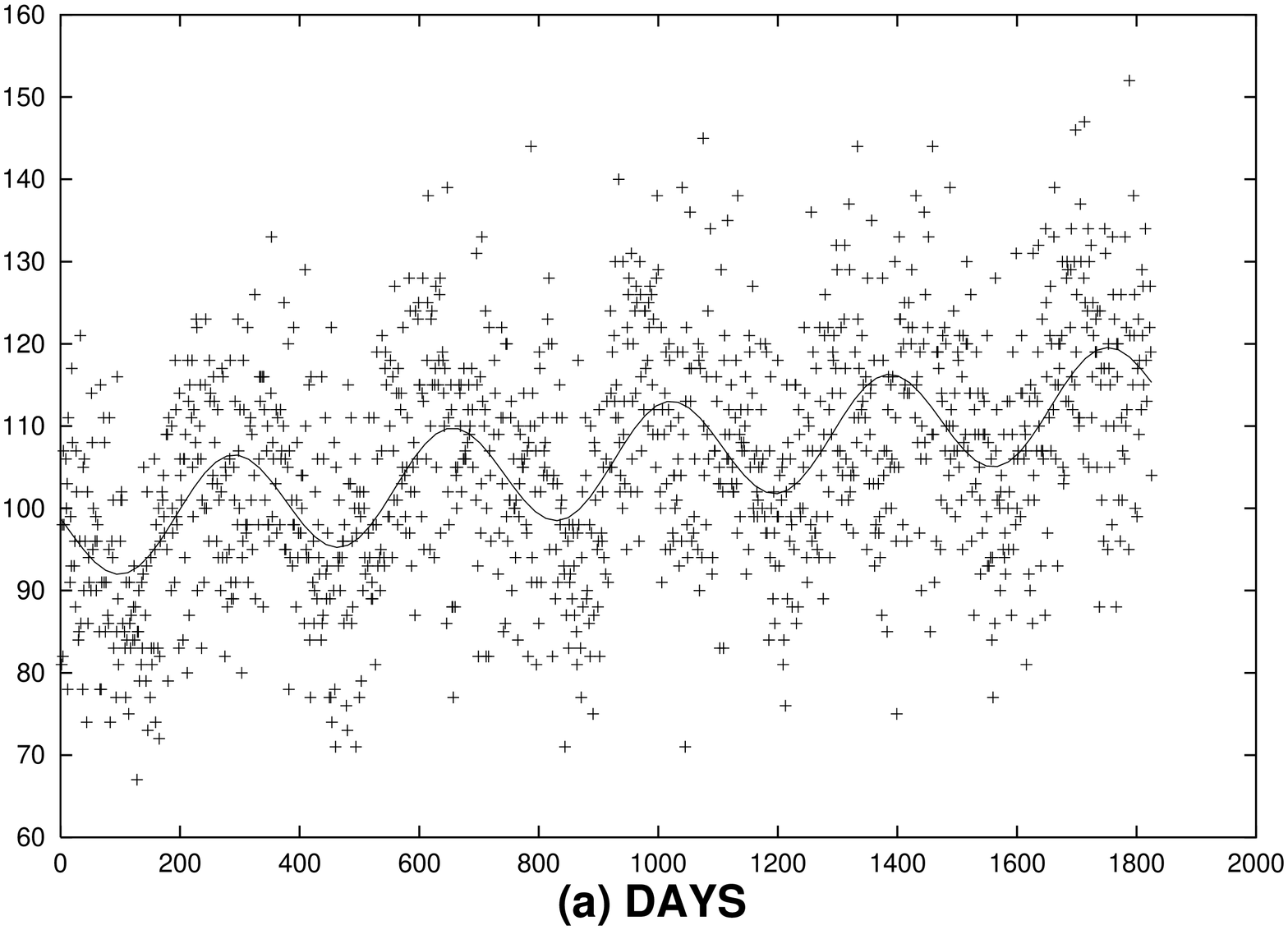, height=16cm,width=8cm,angle=-90}

\epsfig{file=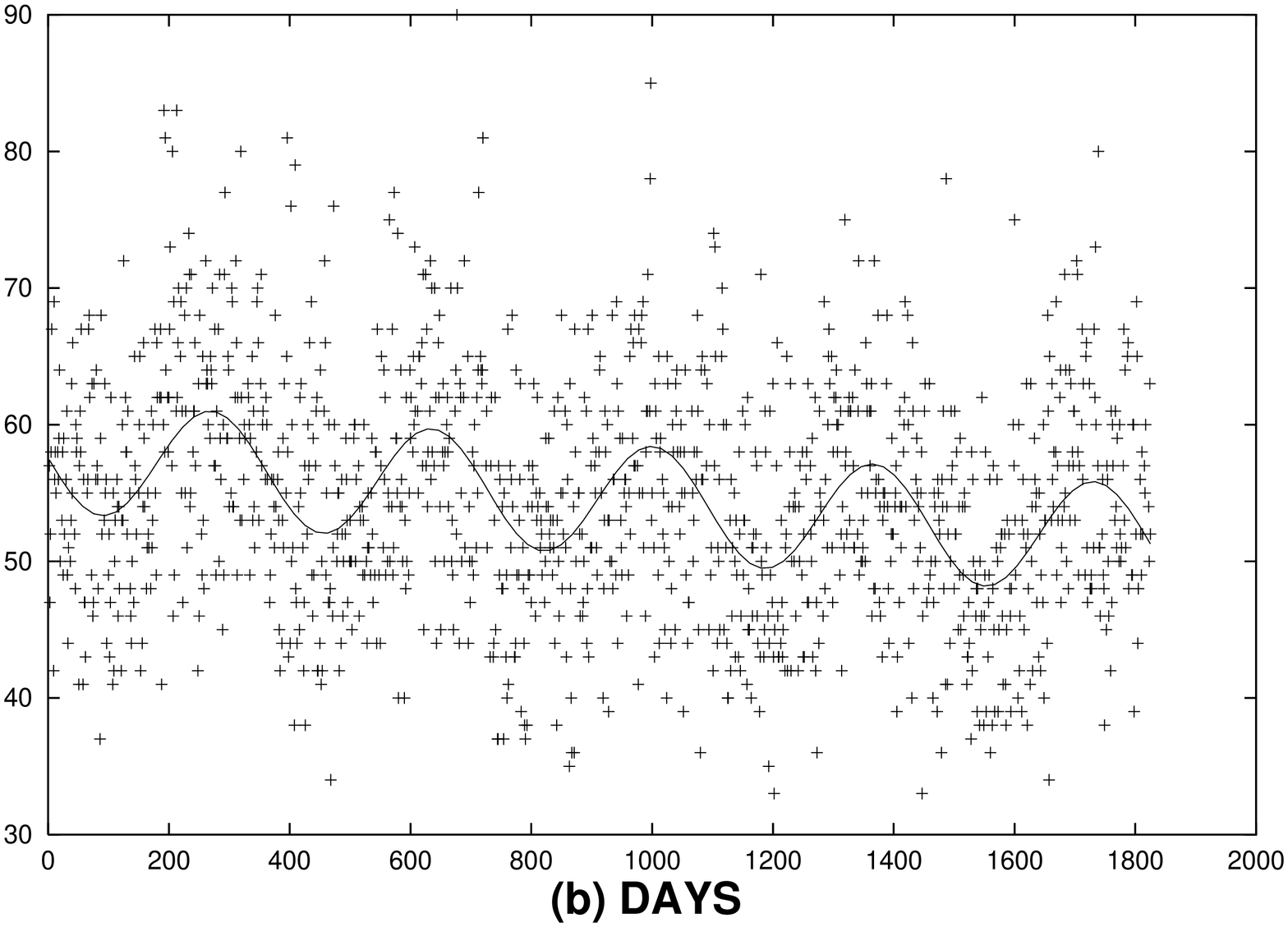, height=16cm,width=8cm,angle=-90}
\caption{(a) The day by day births from unmarried teens in Texas from 
1994 to 1998. (b) The day by day births from married teens in Texas 
from 1994 to 1998. The data has been obtained canceling all the holidays (see the text). The 
solid lines illustrate the choice made to detrend seasonal 
periodicities. This means the analytical expression of Eq. (30) with 
A = 97.5, B = 0.00893, C = 1.29 , D = -6.30 and $\omega = 
2\pi/365.25$, (a), A = 57.8, B = -0.00353, C = -0.227 , D = -4.14 and 
$\omega = 2\pi/365.25$, (b),      }
\end{figure}

\newpage

\begin{figure}[h]
\epsfig{file=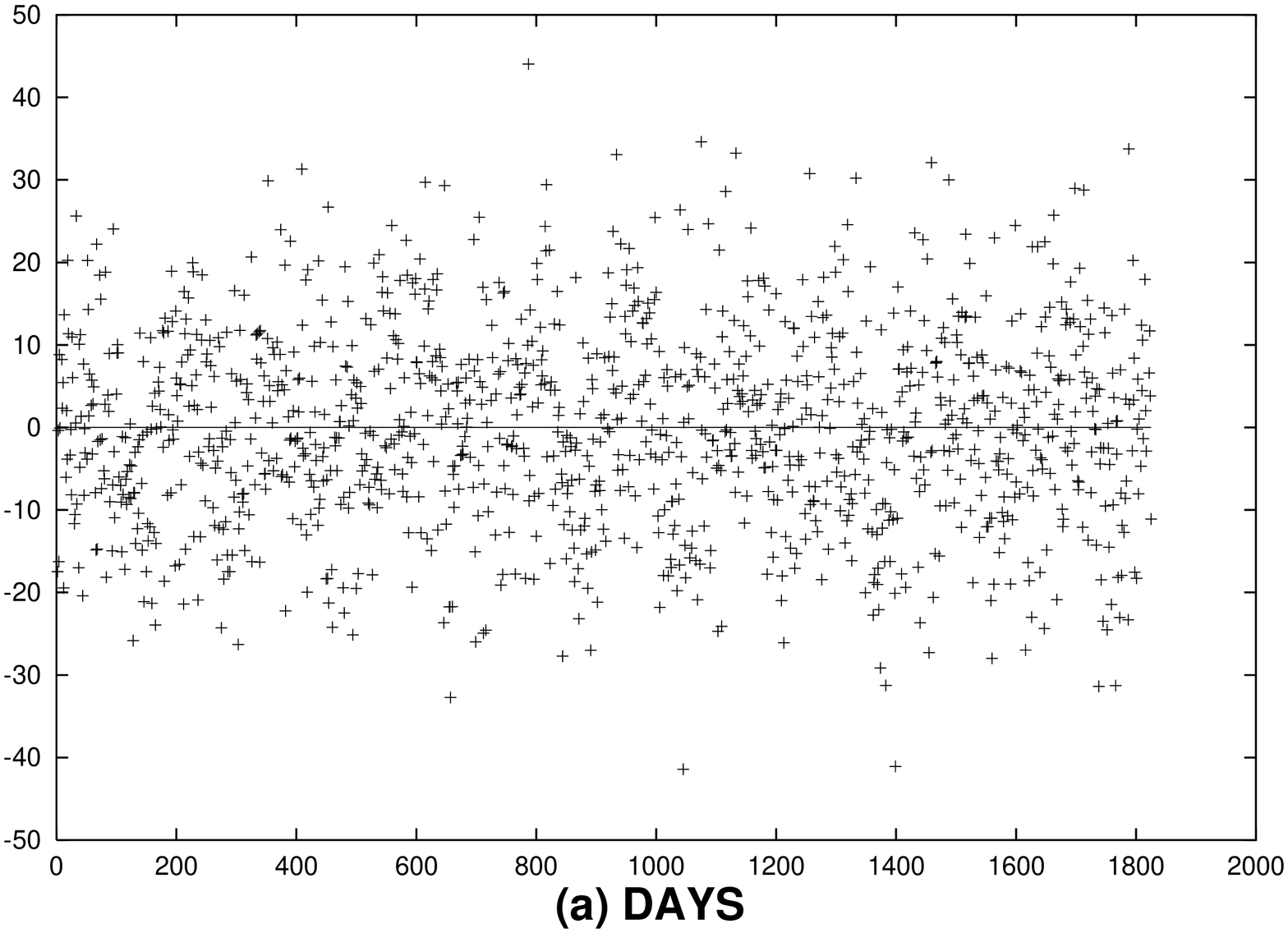, height=16cm,width=8cm,angle=-90}

\epsfig{file=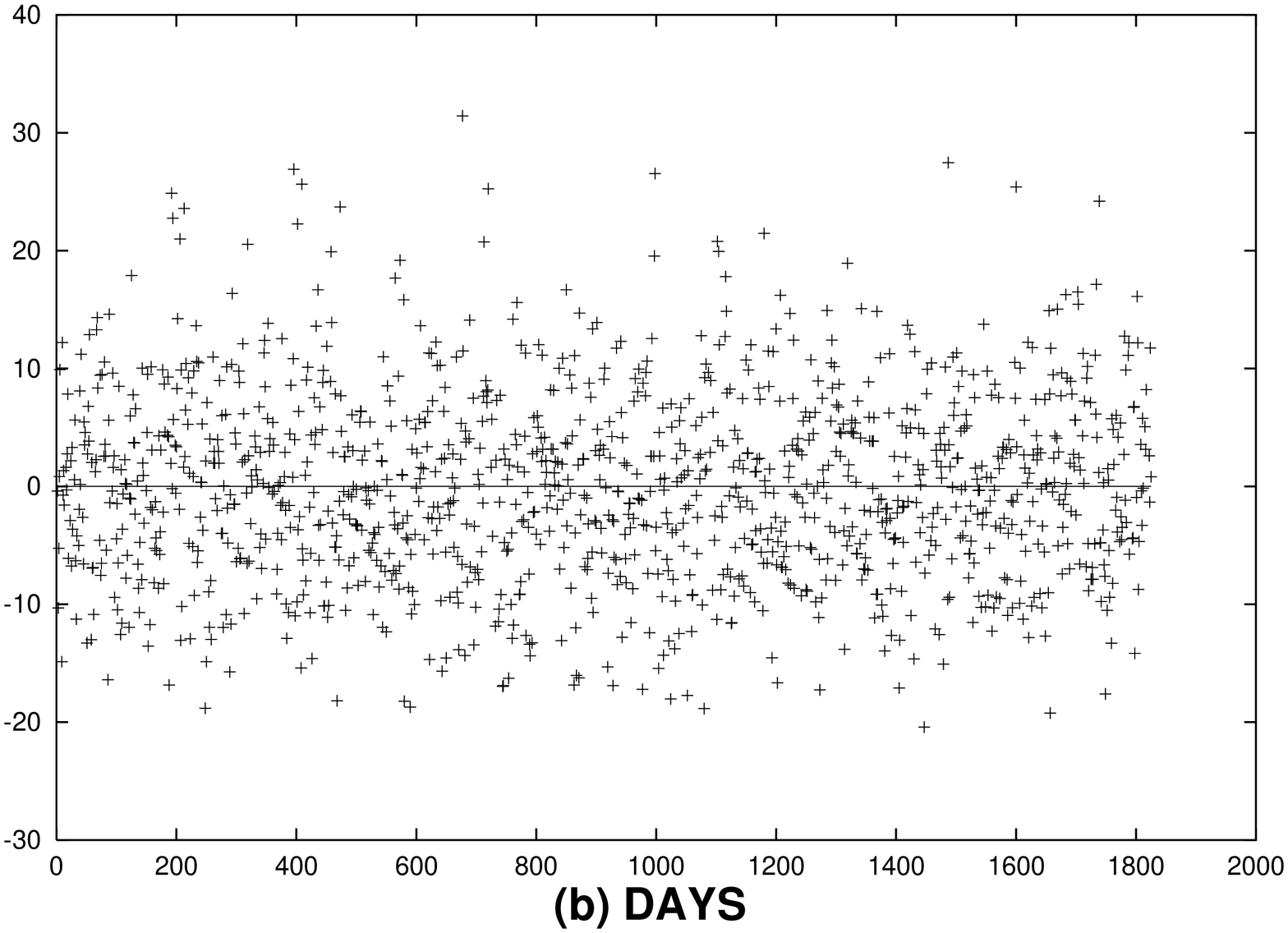, height=16cm,width=8cm,angle=-90}
\caption{(a) and (b) show the data after the
detrending of seasonal periodicity, and of all Saturdays, Sundays and 
holidays respectively for births to unmarried and married teens in 
Texas from 1994 to 1998 . These are the data that we analyze with the diffusion 
entropy method.}
\end{figure}

\newpage

\begin{figure}[h]
\epsfig{file=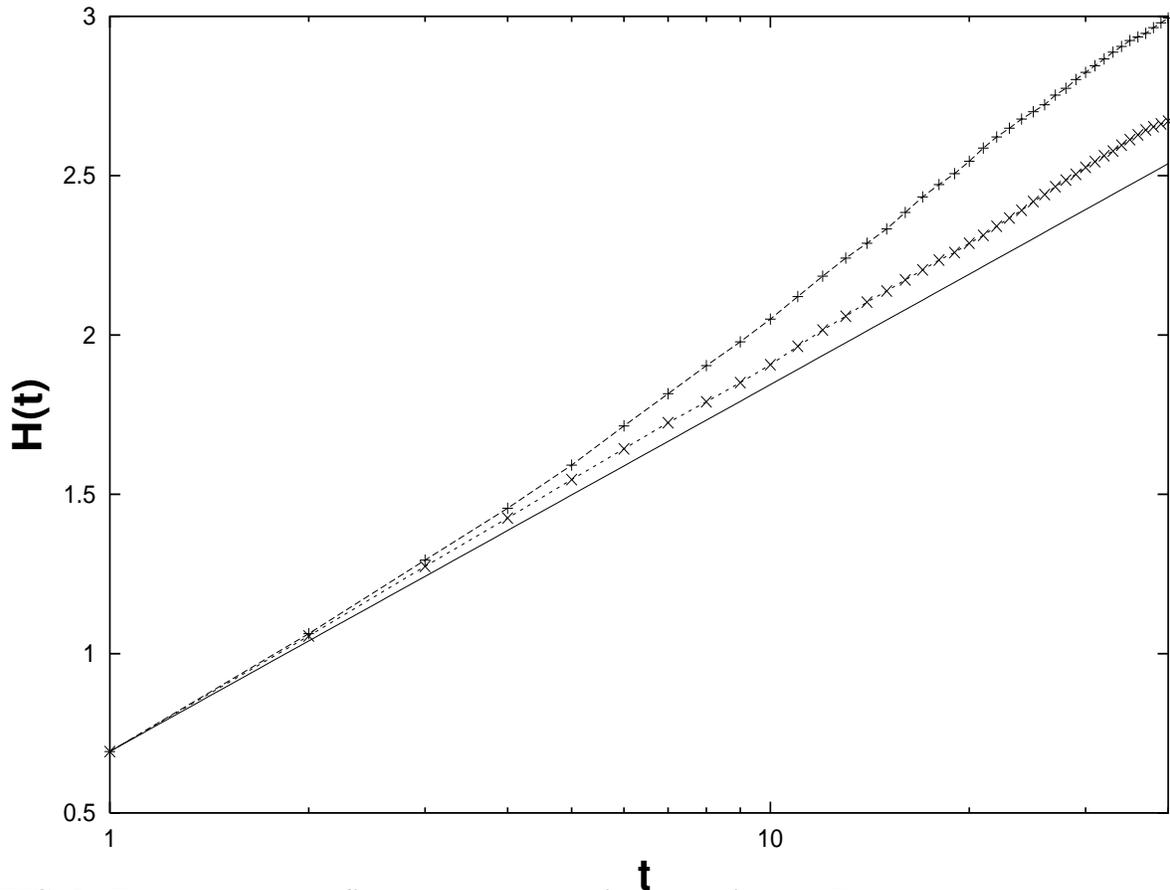, height=16cm,width=12cm,angle=-90}
\caption{The teen births diffusion entropy as a function of time. The 
solid line corresponds
to the prediction of Eq.(\ref{randomwalk}) and serves the main purpose 
of indicating to the reader how the entropy time evolution of a 
stationary process of diffusion would look  in the scale of this 
figure. The case of unmarried teens is denoted by the symbols $+$ and 
the case of married teens is denoted by symbols $\times$. The 
deviation from the straight line of the stationary diffusion process 
of the unmarried teens is stronger than that of the married 
teens.}
\end{figure}

\newpage

\begin{figure}[h]
\epsfig{file=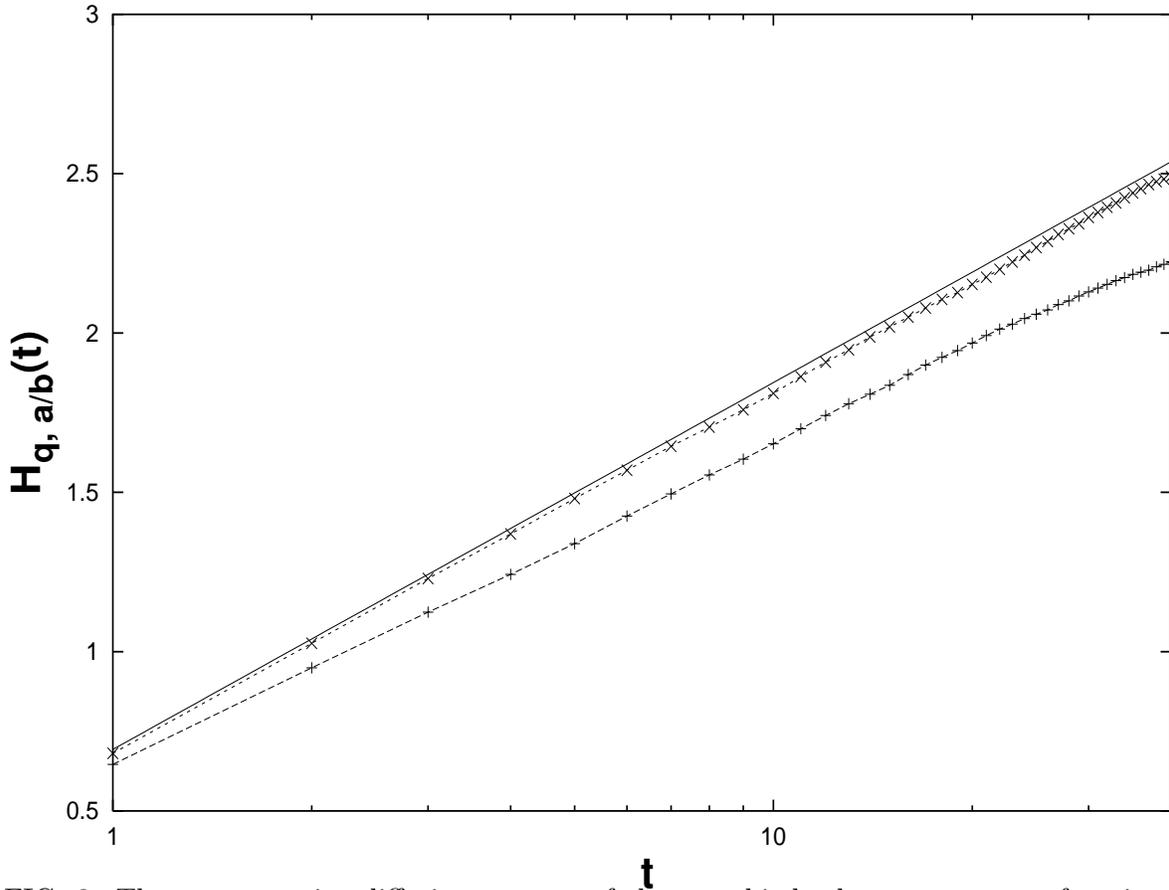, height=16cm,width=12cm,angle=-90}
\caption{The non-extensive diffusion entropy of the teen birth 
phenomenon as a function of time. The solid line refers to the 
prescription of Eq.(\ref{randomwalk}). The case of unmarried and 
married teens are denoted by the symbols $+$ and $\times$, 
respectively. The entropic indices resulting in an entropy increase 
linear with respect to the logarithmic time $\tau$ are q= 1.204 and 
q= 1.050, for unmarried and married teens, respectively.}
\end{figure}

\end{document}